\documentclass[twocolumn,amsthm,secthm]{autart}    

\usepackage{graphicx}          

\usepackage{subfig}
\usepackage{amsmath}
\usepackage{amsfonts}
\usepackage{amssymb}  
\usepackage{bm}
\usepackage{siunitx}
\usepackage{xspace}
\usepackage{xfrac}
\usepackage{empheq}
\usepackage{enumitem}
\usepackage{xcolor}
\usepackage[round]{natbib}

\newcommand{\changed}[1]{#1}
\newcommand{\changedb}[1]{#1}
\DeclareMathOperator{\sign}{sign}
\begin{document}

\begin{frontmatter}

    \title{Worst-Case Error Bounds for the Super-Twisting Differentiator in Presence of Measurement Noise}

\thanks{The financial support by the Christian Doppler Research Association, the Austrian Federal Ministry for Digital and Economic Affairs and the National Foundation for Research, Technology and Development is gratefully acknowledged.
}

\author[IRTCD]{Richard Seeber}\ead{richard.seeber@tugraz.at}

\address[IRTCD]{Christian Doppler Laboratory for Model Based Control of Complex Test Bed Systems, Institute of Automation and Control, Graz University of Technology, Graz, Austria}


\begin{abstract}                          The super-twisting differentiator, also known as the first-order robust exact differentiator, is a well known sliding mode differentiator.
In the absence of measurement noise, it achieves exact reconstruction of the time derivative of a function with bounded second derivative.
This note proposes an upper bound for its worst-case differentiation error in the presence of bounded measurement noise, based on a novel \changed{Lipschitz continuous} Lyapunov function.
It is shown that the bound can be made arbitrarily tight and never exceeds the true worst-case differentiation error by more than a factor of two.
A numerical simulation \changed{illustrates the results and also} demonstrates the non-conservativeness of the proposed bound.
 \end{abstract}

\end{frontmatter}

\begingroup
\def\QED{$\blacksquare$}
\newcommand{\abs}[1]{\left\lvert #1 \right\rvert}
\newcommand{\norm}[2][]{\left\lVert #2 \right\rVert_{#1}}

\newcommand{\diffd}{\mathrm{d}}
\newcommand{\dt}[1]{\deriv{#1}{t}}

\newcommand{\ceil}[1]{\left\lceil{#1}\right\rceil}
\newcommand{\floor}[1]{\left\lfloor{#1}\right\rfloor}

\newcommand{\spow}[2]{\left\lfloor #1 \right\rceil^{#2}}
\newcommand{\spowf}[3]{\spow{#1}{\frac{#2}{#3}}}
\newcommand{\apow}[2]{\abs{#1}^{#2}}
\newcommand{\apowf}[3]{\apow{#1}{\frac{#2}{#3}}}

\newcommand{\sbrm}[1]{\sb{\mathrm{#1}}}

\newcommand{\TT}{^{\mathrm{T}}}
\newcommand{\HH}{^{\mathrm{H}}}

\newcommand{\RR}{\mathbb{R}}
\newcommand{\CC}{\mathbb{C}}

\def\x{\mathbf{x}}
\let\oldepsilon\epsilon
\let\epsilon\varepsilon
 \newcommand{\FL}{\mathcal{F}_L}
\newcommand{\EN}{\mathcal{E}_N}

\section{Introduction}

Differentiation of noisy signals is of importance in many engineering applications.
Examples include velocity estimation, fault detection, and active vibration damping.
Hence, differentiators have been studied extensively in literature, see e.g. the special issue by \cite{reichhartinger2018special}, with linear high gain differentiators \citep{vasiljevic2008error,khapra_ijrnc14}, algebraic differentiators \citep{mbojoi_na09,mboup2018frequency}, and robust exact differentiators based on sliding modes \citep{levant_ijc03,levliv_ejc20,moreno2022arbitrary} being some popular approaches.
The latter, in particular, have the attractive feature that they differentiate exactly in the absence of measurement noise, and that they can be tuned without knowledge about the noise.

All differentiators are impacted to some extent by measurement noise.
For linear high-gain differentiators, \cite{vasiljevic2008error} derive upper bounds on the differentiation error and show how to optimally tune their parameters if the amplitude of the noise is known.
For first-order differentiation, such an optimal tuning yields a differentiation error upper bound of the form $c \sqrt{N L}$ with a constant $c \ge 2$, \changed{cf. \cite[Proposition 3.1]{seeber2023optimal}}, where $N$ is a bound for the measurement noise $\eta$, and $L$ is a bound for the second derivative of the function $f$ whose first-order derivative is to be computed from the noisy measurement $u = f + \eta$.
The first-order robust exact differentiator by \cite{levant1998robust}, also known as the super-twisting differentiator, also exhibits a differentiation error upper bound of the form $c \sqrt{N L}$ with\footnote{\changed{\emph{Non-causal} exact differentiators can achieve smaller values for $c$; in that case, $c \ge 2$ as shown by \cite{levant2017sliding}.}} $c \ge 2 \sqrt{2}$, \changed{cf. \cite[Proposition 3.3]{seeber2023optimal}}, but without requiring the knowledge of $N$.
This is shown qualitatively by \cite{levant1998robust}, but only very few works, by \cite{angulo2012differentiation,angulo2012optimal}, tackle the quantitative computation of the bound.
When applied correctly, their approach yields very conservative bounds, however.
Alternatively, Lyapunov functions such as in \citep{cruz2019levant} may be used.
Bounds resulting from those are only given in examples, however, and to the best of the author's knowledge, no rigorous quantitative upper bounds for the differentiation error of the super-twisting differentiator exist in literature today.\footnote{Some other works, such as \citep{efimov2011hybrid}, compute such bounds for other sliding mode differentiators.}

This note proposes a non-conservative upper bound for the differentiation error of the super-twisting differentiator in the presence of noisy measurements.
In particular, it is shown that the proposed bound can be made arbitrarily tight and is always within a factor of two of the actual worst-case error.
Moreover, the tuning of the differentiator based on a tradeoff between the convergence speed and the differentiation error is discussed.

Section~\ref{sec:prelim} introduces the super-twisting differentiator and the problem of signal differentiation from noisy measurements.
Section~\ref{sec:main} then presents the main results: a lower and---in main Theorem~\ref{thm:main}---an upper bound for the worst-case differentiation error, along with a guideline for parameter tuning.
Section~\ref{sec:lyap} proposes a novel \changed{Lipschitz continuous} Lyapunov function that is used to formally prove the proposed upper bound.
Section~\ref{sec:sim}, finally, illustrates the proposed bound in a numerical simulation.
Section~\ref{sec:conclusion} draws conclusions, and an appendix contains proofs of a proposition and two lemmata.

\textbf{Notation:}
Vectors are written as boldface lowercase letters, and $\RR$, $\RR_{\ge 0}$, and $\RR_{>0}$ denote the reals, nonnegative reals, and positive reals, respectively.
For $y,p \in \RR$, $|y|$ is the absolute value of $y$, $\lfloor y \rfloor$ is the largest integer not greater than $y$, and the abbreviation $\spow{y}{p} = \apow{y}{p} \sign(y)$ is used; in particuar, $\spow{y}{0} = \sign(y)$.
\changed{The convex hull of a set $A \subseteq \RR$ is denoted by $\operatorname{co} A$.}

\section{Super-twisting differentiator}
\label{sec:prelim}

Consider the super-twisting differentiator, also known as Levant's first-order robust exact differentiator \citep{levant1998robust}, given by the system
\begin{subequations}
    \label{eq:diff}
    \begin{align}
        \dot y_1 &= \lambda_1 \sqrt{L} \spowf{u - y_1}{1}{2} + y_2, & y_1(0) &= u(0), \\
        \dot y_2 &= \lambda_2 L \spow{u-y_1}{0}, & y_2(0) &= 0, \\
y &= y_2
    \end{align}
\end{subequations}
with input $u$, output $y$, and positive parameters $\lambda_1$, $\lambda_2$, and $L$.
Solutions of this system with discontinuous right-hand side are understood in the sense of \cite{filippov1988differential}.
The input $u = f + \eta$ consists of a signal \changed{$f : \RR_{\ge 0} \to \RR$} to be differentiated with second derivative bounded by $|\ddot f(t)| \le L$ almost everywhere, and a measurement noise \changed{$\eta : \RR_{\ge 0} \to \RR$} bounded by $|\eta(t)| \le N$ for all $t \ge 0$.
The sets of admissible signal and noise functions are given by
	\begin{align}
		\FL &= \{ f \in \mathcal{F} : |\ddot f(t)| \le L \text{ almost everywhere on $\RR_{\ge 0}$} \}, \nonumber \\
\EN &= \{ \eta \in \mathcal{E} : |\eta(t)| \le N \text{ on $\RR_{\ge 0}$} \},
	\end{align}
wherein $\mathcal{F}$ denotes the set of differentiable functions with Lipschitz continuous first derivative and $\mathcal{E}$ denotes the set of Lebesgue measurable functions.
It is well known that after a finite time $\tau$ depending only on\footnote{The initial condition $y_1(0) = u(0)$ ensures that the convergence time $\tau$ is independent of $f(0)$.
This is useful in practice and will become obvious from the error system \eqref{eq:differror} later on.} \changed{$\dot f(0)$ as well as $\lambda_1, \lambda_2$, and $L,N$}, the differentiation error is bounded by 
$
|y(t) - \dot f(t)| \le c \sqrt{NL}
    $
for all $t \ge \tau$ with some constant $c$ depending only on $\lambda_1$ and $\lambda_2$, cf. \citep{levant_ijc03}.

\section{Main Results}
\label{sec:main}

\changed{This section shows lower and upper bounds for the worst-case differentiation error of the differentiator~\eqref{eq:diff}.}

\subsection{Worst-case differentiation error lower bound}

To put the \changedb{results} into context, a lower bound on the worst-case differentiation error is stated first.
For every selection of parameters, existence of some worst-case signals $f$ and $\eta$ is shown that lead to an error of at least $2\sqrt{\lambda_2+1}\sqrt{NL}$, \changedb{analogous to \cite[Proposition 3.3]{seeber2023optimal} but without requiring differentiator convergence.}
The proof is given in Appendix~\ref{app:proofs:lowerbound}.
\begin{prop}
    \label{prop:lowerbound}
    Let \changed{$L,N \in \RR_{> 0}$}. Consider differentiator \eqref{eq:diff} with parameters $\lambda_1, \lambda_2 \in \RR_{> 0}$.
Then, for every $\tau \in \RR_{\ge 0}$ there exist $f \in \FL$, $\eta \in \EN$ with \changed{$f(0) = \dot f(0) = 0$} such that a trajectory of \eqref{eq:diff} with input $u = f + \eta$ satisfies
\vspace{-0.2em}\begin{equation}
        \label{eq:lowerbound}
        \sup_{t \ge \tau} \abs{y(t) - \dot f(t)} \ge 2\sqrt{\lambda_2 +1} \sqrt{N L}.
    \end{equation}
\end{prop}

\subsection{Worst-case differentiation error upper bound}

The following main theorem, proven in Section~\ref{sec:proof:main}, establishes an upper bound for the differentiation error.
\begin{thm}
    \label{thm:main}
    Let $L \in \RR_{> 0}$, $N \in \RR_{\ge 0}$, and $\alpha \in (1,4]$. Consider the differentiator \eqref{eq:diff} and suppose that its parameters $\lambda_1, \lambda_2$ satisfy
\vspace{-0.2em}\begin{equation}
        \label{eq:cond}
1 < \frac{\lambda_1}{\sqrt{8 (\lambda_2 + 1)}} < \frac{(\alpha + 1) \lambda_2 + \alpha - 1}{2 \sqrt{\alpha} (\lambda_2 +1)}.
    \end{equation}
\vspace{-0.2em}Then, for \changedb{each} input $u = f + \eta$ with $f \in \FL, \eta \in \EN$, there exists a finite time $\tau$ depending only on \changed{the corresponding initial values} $\dot f(0)$ and $\eta(0)$, such that \changed{for all $t \ge \tau$}
\begin{equation}
        \label{eq:accuracy}
        \abs{y(t) - \dot f(t)} \le 2 \sqrt{\alpha (\lambda_2 +1 )} \sqrt{NL}.
    \end{equation}
\end{thm}
\begin{rem}[Parameter range]
    \label{rem:range}
    Note that a value for $\lambda_1$ satisfying \eqref{eq:cond} exists if and only if
    \begin{equation}
        \label{eq:lambda2min}
        \lambda_2 > \frac{1 + 2 \sqrt{\alpha} - \alpha}{1- 2 \sqrt{\alpha} +\alpha} \ge 1
    \end{equation}
    with $\alpha \in (1,4]$ as in the theorem.
    This condition is least restrictive for $\alpha = 4$, in which case a $\lambda_1$ may be selected for every $\lambda_2 > 1$ according to
    \begin{equation}
        \frac{\lambda_1}{\sqrt{8 (\lambda_2 + 1)}} \in \left( 1,  1 + \frac{\lambda_2 - 1}{4(\lambda_2 + 1)}\right).
\end{equation}
\end{rem}

\begin{rem}[Tightness of the bound]
    By comparing \eqref{eq:accuracy} to \eqref{eq:lowerbound}, one can see that the proposed upper bound \changed{never exceeds the actual worst-case differentiation error by more than a factor of $\sqrt{\alpha} \le 2$}, and that it can be made arbitrarily tight by reducing $\alpha \in (1,4]$.
This restricts $\lambda_2$ according to Remark~\ref{rem:range}, however, thus possibly increasing the worst-case differentiation error, which may be undesirable in practice.
    Nevertheless, the simulation results in Section~\ref{sec:sim} show that non-conservative bounds may also be obtained for values of $\alpha$ as large as $\alpha = 4$.
\end{rem}

\begin{rem}[Tuning guideline]
    \label{rem:tuning}
    Subject to the parameter condition \eqref{eq:cond} and with the initial condition as in~\eqref{eq:diff}, 
\cite[Theorem 3.15]{seeber2020computing}
    shows that, in the absence of noise, the super-twisting differentiator's output $y$ converges to the true derivative $\dot f$ of $f$ after the time $\frac{|\dot f(0)|}{(\lambda_2 - 1) L}$ in the worst case.
    Using this expression along with \eqref{eq:accuracy}, a quantitative tradeoff between noise-free worst-case convergence time and worst-case differentiation error in the presence of noise may be found by noting that increasing $\lambda_2$ decreases the former and increases the latter.
\end{rem}

\section{Lyapunov Function and Proof of the Bound}
\label{sec:lyap}

To prove the worst-case error upper bound, a novel Lyapunov function for the super-twisting differentiator is proposed.
To that end, the error states $x_1 = y_1 - f$, $x_2 = y_2 - \dot f$ are introduced and aggregated in the vector $\x = [ x_1 \quad x_2]\TT$.
They are governed by the dynamics
\begin{subequations}
    \label{eq:differror}
    \begin{align}
        \dot x_1 &= -\lambda_1 \sqrt{L} \spowf{x_1 - \eta}{1}{2} + x_2, & x_1(0) &= \eta(0), \\
        \dot x_2 &= -\lambda_2 L \spow{x_1 - \eta}{0} - \ddot f, & x_2(0) &=  -\dot f(0).
    \end{align}
\end{subequations}
\subsection{Lyapunov function}

The proposed Lyapunov function $V : \RR^2 \to \RR$ is
\begin{equation}
    \label{eq:V}
    V(\x) = \begin{cases}
        W(\x) & \text{if } x_2 \ge 0 \\
        W(-\x) & \text{if } x_2 < 0
    \end{cases}
\end{equation}
with $W : \RR \times \RR_{\ge 0} \to \RR$ defined as
\begin{equation}
    \label{eq:W}
    W(\x) = \begin{cases}
        W_1(\x) & \text{if } x_1 \le \frac{x_2^2}{4 \alpha (\lambda_2 + 1) L} \\
        W_2(\x) & \text{if } \frac{x_2^2}{4 \alpha (\lambda_2 + 1) L} < x_1 \le \frac{(2\alpha + 1) x_2^2}{4\alpha (\lambda_2 + 1) L} \\
        W_3(\x) & \text{if } \frac{(2\alpha +1) x_2^2}{4\alpha (\lambda_2 + 1) L} < x_1,
    \end{cases}
\end{equation}
wherein the abbreviations
\begin{subequations}
    \label{eq:Wi}
    \begin{align}
        W_1(\x) &= \frac{x_2^2}{2 \alpha(\lambda_2 + 1) L} - x_1, \displaybreak[0]\\
        W_2(\x) &= \frac{x_2^2}{4 \alpha (\lambda_2 + 1) L}, \\
        W_3(\x) &= x_1 - \frac{x_2^2}{2 (\lambda_2 + 1) L}
    \end{align}
\end{subequations}
are used and $\alpha \in (1,4]$ is a constant parameter.
Fig.~\ref{fig:contour} shows the level curves of this Lyapunov function for parameter values $\lambda_2 = 1.1$, $L = 1$, $\alpha = \frac{4}{2.1}$.
\begin{figure}[tbp]
    \centering
    \includegraphics{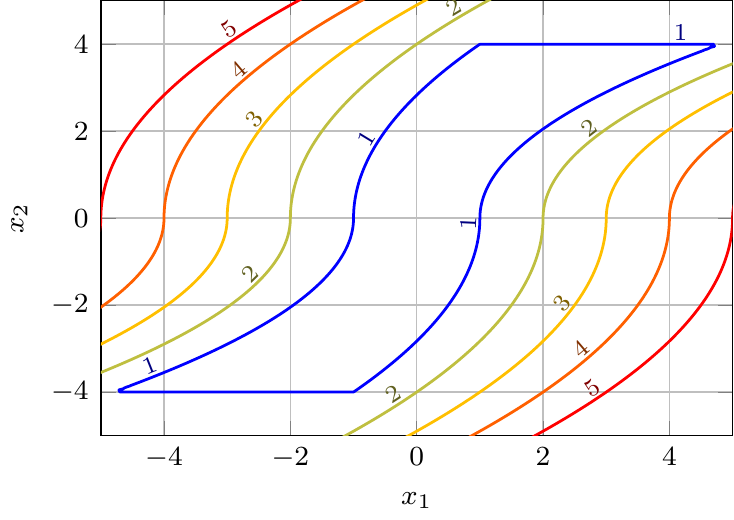}
    \caption{Level curves and values of the Lyapunov function \eqref{eq:V}--\eqref{eq:Wi} for parameter values $\lambda_2 = 1.1$, $L = 1$, $\alpha = \frac{4}{2.1}$.}
    \label{fig:contour}
\end{figure}

\subsection{Auxiliary lemmata}

The main result is proven by means of a differential inequality for the Lyapunov function $V$ that is established in the following lemma.
It is proven in Appendix~\ref{app:proofs:lyap}.
\begin{lem}
\label{lem:lyap}
Let $L \in \RR_{>0}$, $N \in \RR_{\ge 0}$, $\alpha \in (1,4]$, and suppose that $\lambda_1, \lambda_2$ satisfy \eqref{eq:cond}.
Consider $V : \RR^2 \to \RR$ as defined in \eqref{eq:V}--\eqref{eq:Wi}.
Then, $V$ is \changed{Lipschitz} continuous and positive definite, and there exists $\gamma > 0$ such that the time derivative $\dot V$ of $V$ along the trajectories of \eqref{eq:differror} \changed{with $|\ddot f| \le L$, $|\eta|\le N$} fulfills\footnote{\changed{Along a trajectory $\x(t)$, \eqref{eq:Vdot} holds almost everywhere in time $t$, cf. \cite[Section 5.4]{polyakov2014stability}.}}
\begin{equation}
    \label{eq:Vdot}
    \dot V(\x) \le - \gamma \sqrt{V(\x) - N}
\end{equation}
for all $\x \changed{\in \RR^2}$ with $V(\x) > N$.
\end{lem}

Using this result, the set characterized by $V(\x) \le N$ is shown to be forward invariant and attractive in finite time.
The following lemma establishes this, along with an upper bound for $|x_2|$.
It is proven in Appendix~\ref{app:proofs:inv}.
\begin{lem}
    \label{lem:inv}
    Let $L \in \RR_{>0}$, $N \in \RR_{\ge 0}$, $\alpha \in (1,4]$, and $\lambda_1, \lambda_2 \in \RR_{> 0}$.
Consider the function $V : \RR^2 \to \RR$ as defined in \eqref{eq:V}--\eqref{eq:Wi} and define the set
    $
        \Omega = \{ \x \in \RR^2 : V(\x) \le N \}.
        $
    Then,
    \begin{equation}
        \label{eq:supV}
        \sup_{\x \in \Omega} \abs{x_2} = 2\sqrt{\alpha (\lambda_2 + 1) N L}.
    \end{equation}
    Moreover, if $\lambda_1, \lambda_2$ satisfy \eqref{eq:cond}, then for every trajectory $\x : \changed{\RR_{\ge 0}} \to \RR^2$ of \eqref{eq:differror} \changed{with $|\ddot f| \le L$, $|\eta|\le N$} there exists a finite time $\tau$ depending only on $\x(0)$ such that $ \x(t) \in \Omega$ holds for all $t \ge \tau$.
\end{lem}

Fig.~\ref{fig:contour} depicts the set $\Omega$ for $N = 1$ as the interior of the level curve characterized by $V(\x) = 1$.
With the particular parameters $\lambda_2 = 1.1$, $L = 1$, $\alpha = \frac{4}{2.1}$ of the depicted Lyapunov function, the (tight) inequality $\abs{x_2} \le 2 \sqrt{\alpha (\lambda_2 + 1) NL} = 4$ can be seen to hold in $\Omega$.

\subsection{Proof of Theorem~\ref{thm:main}}
\label{sec:proof:main}

Theorem~\ref{thm:main} may now be proven.
Consider the error system \eqref{eq:differror}.
From Lemma~\ref{lem:inv}, there exists a finite time $\tau$ depending on $\x(0) = [\eta(0) \quad -\dot f(0)]\TT$ such that $\x(t) \in \Omega$ for all $t \ge \tau$.
The claim follows from the fact that
\begin{equation}
    \abs{y(t) - \dot f(t)} = \abs{x_2(t)} \le \sup_{\x \in \Omega} \abs{x_2} = 2 \sqrt{\alpha (\lambda_2 + 1) N L}
\end{equation}
for $\x(t) \in \Omega$ according to \eqref{eq:supV}.\qed

\section{Simulation Results}
\label{sec:sim}

This section \changed{illustrates} the obtained worst-case error upper bound \eqref{eq:accuracy} \changed{by means of} a numerical simulation.
\changed{To that end, the parameters $L = 1$, $\lambda_2 = 1.1$ are considered and, to satisfy \eqref{eq:cond}, $\alpha = 4$ and $\lambda_1 = 4.1$ are chosen.}
The simulation is performed using an implicit discretization of the super-twisting differentiator \eqref{eq:diff} as in \cite{mojallizadeh2021time}, with sampling time step $\Delta = 5 \cdot 10^{-4}$.
\changed{The differentiator is applied to $f(t) = -L t^2/2$ and}
\begin{equation}
    \label{eq:etasim}
    \eta(t) = \begin{cases}
        -N & \text{if } t < 10 c_1 \\
        -N \spow{t - c_1 \left\lfloor \frac{t}{c_1} \right\rfloor - c_2}{0} & \text{if } t \ge 10 c_1 \\
    \end{cases}
\end{equation}
with parameters $N = 0.01$ and $c_1 = 0.011$, $c_2 = 0.00149$ is used, corresponding to a high-frequency noise with duty cycle $c_2/c_1 \changed{\approx \SI{13.5}{\percent}}$.
Fig.~\ref{fig:simulation} depicts the resulting differentiation error along with the proposed upper bound \eqref{eq:accuracy}, as well as the noise signal.
One can see that the high frequency noise leads to an error close to the proven upper bound, demonstrating the non-conservativeness of the bound.

\begin{figure}[tbp]
    \centering
    \includegraphics{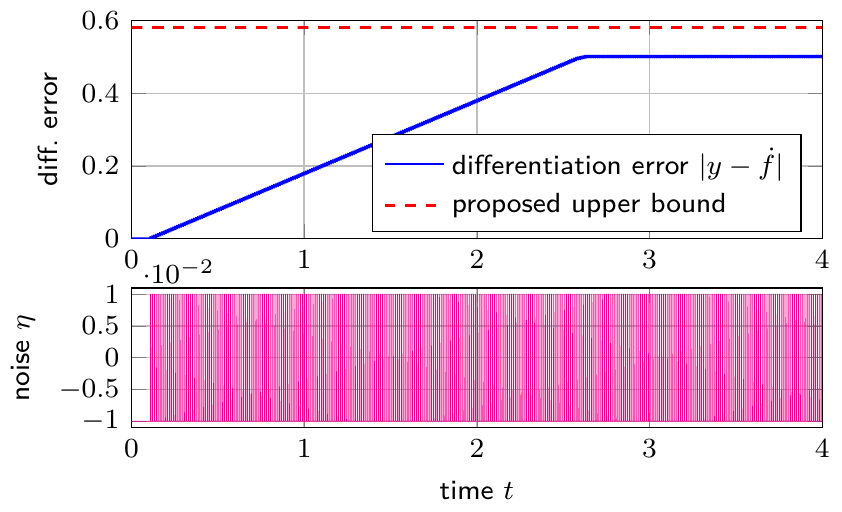}
    \caption{Differentiation error from a numerical simulation of an implicit discretization of \eqref{eq:diff} with sampling time step $\Delta = 5\cdot 10^{-4}$ and parameters $\lambda_1 = 4.1$, $\lambda_2 = 1.1$, $L = 1$ differentiating the signal $f(t) = L t^2/2$ with measurement noise $\eta(t)$ as in \eqref{eq:etasim} with $N = 0.01$, $c_1 = 0.011$, $c_2 = 0.00149$.
    Also shown is the proposed differentiation error upper bound \eqref{eq:accuracy} with $\alpha = 4$, given by $2 \sqrt{\alpha (\lambda_2 + 1) N L} \approx 0.58$.}
    \label{fig:simulation}
\end{figure}

\section{Conclusion and Outlook}
\label{sec:conclusion}
Based on a novel Lyapunov function for the super-twisting differentiator, a non-conservative upper bound for its differentiation error in presence of noisy measurements was proposed.
The simple form of the proposed bound makes it easy to use for the purpose of finding a quantitative tradeoff between convergence speed and differentiation error when tuning the differentiator in practice.
Moreover, the proposed bound never exceeds the true worst-case error by more than a factor of two and can also be made arbitrarily tight.
Future work may study possible extensions of the obtained results to 
discretized variants of the super-twisting differentiator.

\appendix

\section{Proofs}
\label{app:proofs}

\subsection{Proof of Proposition~\ref{prop:lowerbound}}
\label{app:proofs:lowerbound}

If $\lambda_2 < 1$, then choosing $u = f+ \eta$ with $f(t) = L t^2/2$, $\eta(t) = N$ in \eqref{eq:diff} yields $y(t) \le \lambda_2 L t$, and hence the error $y(t) - \dot f(t) \le (\lambda_2 - 1) L t$ is unbounded.
For $\lambda_2 \ge 1$, let
$
    \theta = 2 \sqrt{\frac{N}{(\lambda_2+1) L}},
    $
let $\tau > \theta$ without restricting generality, and consider
\begin{align}
    f(t) &= \begin{cases}
        0 &  \text{if } t < \tau - \theta \\
        \frac{L (t-\tau + \theta)^2}{2} & \text{if } t \ge \tau - \theta,
    \end{cases} \\
    \eta(t) &= \max\{ -N, N - (\lambda_2 + 1) f(t) \}.
\end{align}
Then, $y_1(t) = u(t) = N - \lambda_2 f(t)$, $y_2(t) = \dot u(t)$ is a sliding mode trajectory of \eqref{eq:diff} for $t \le \tau$, yielding
\begin{equation}
    y(\tau) - \dot f(\tau) = -(\lambda_2 + 1) \dot f(\tau) = - 2 \sqrt{\lambda_2 + 1} \sqrt{N L},
\end{equation}
thus proving the claim.\qed

\subsection{Proof of Lemma~\ref{lem:lyap}}
\label{app:proofs:lyap}

\changed{Lipschitz} continuity is straightforward to verify.
Positive definiteness follows from the fact that $W(\x) < 0$ yields a contradiction in each case of \eqref{eq:W}.
To show \eqref{eq:Vdot}, the time derivative of $V$ is computed for each of the three cases in \eqref{eq:W}.
All computations are performed supposing that $V > N$ and $x_2 \ge 0$ hold; the results are then equally valid for $x_2 < 0$ due to symmetry.
In doing so, it is worth noting that \eqref{eq:cond} implies $\lambda_2 > 1$ according to Remark~\ref{rem:tuning}.

In the first case in \eqref{eq:W}, $\dot V = \dot W_1$ holds with
\begin{align}
    \dot W_1 &= \frac{x_2}{\alpha (\lambda_2 + 1) L} (-\lambda_2 L \spow{x_1 -\eta}{0} - \ddot f) \nonumber \\
    &\quad+ \lambda_1 \sqrt{L} \spowf{x_1 - \eta}{1}{2} - x_2 \nonumber \\
    \label{eq:W1dot}
    &\le \frac{-\lambda_2 \spow{x_1 -\eta}{0} + 1}{\alpha(\lambda_2 + 1)} x_2 + \lambda_1 \sqrt{L} \spowf{x_1 - \eta}{1}{2} - x_2.
\end{align}
For $x_1 - \eta \le 0$, one obtains
\begin{equation*}
    \dot W_1 \le -\frac{\alpha -1}{\alpha} x_2 + \lambda_1 \sqrt{L} \spowf{x_1 - \eta}{1}{2}.
\end{equation*}
If, additionally, $x_2 \le \sqrt{\alpha (\lambda_2 +1) L} \sqrt{V -N}$, then
\begin{equation}
x_1 - \eta = \frac{x_2^2}{2 \alpha (\lambda_2 + 1) L} - V - \eta \le -\frac{V-N}{2}
\end{equation}
by definition of $V$; hence 
$
    \dot V = \dot W_1 \le - \lambda_1 \sqrt{L/2} \sqrt{V-N}.
$
Otherwise, if $x_2 > \sqrt{\alpha (\lambda_2 +1) L} \sqrt{V -N}$, then 
\begin{equation}
    \dot V = \dot W_1 \le -\frac{\alpha -1}{\alpha} \sqrt{\alpha (\lambda_2 +1) L} \sqrt{V -N}.
\end{equation}
For $x_1 - \eta > 0$, one obtains
\begin{align}
    \dot W_1 &\le \left( \frac{-\lambda_2 + 1}{\alpha (\lambda_2 +1)} - 1\right) x_2 + \lambda_1 \sqrt{L} \spowf{x_1 - \eta}{1}{2} \nonumber \\
    \label{eq:W1dot2}
    &= - \frac{(\alpha + 1) \lambda_2 + \alpha - 1}{\alpha (\lambda_2 + 1)} x_2 + \lambda_1 \sqrt{L} \spowf{x_1 - \eta}{1}{2}
\end{align}
from \eqref{eq:W1dot}.
By definition of $V$, one moreover has
\begin{align}
    \label{eq:x1eta1}
    0 < x_1 - \eta &\le \frac{x_2^2}{2 \alpha (\lambda_2 + 1) L} - V -\eta \le \frac{x_2^2}{2 \alpha (\lambda_2 + 1) L}.
\end{align}
\changed{since $-\eta < V$.}
Substituting for $x_1 - \eta$ in \eqref{eq:W1dot2} yields
\begin{equation*}
    \dot W_1 \le - \frac{(\alpha + 1) \lambda_2 + \alpha - 1}{\alpha (\lambda_2 + 1)} x_2 + \frac{\lambda_1}{\sqrt{2 \alpha (\lambda_2 + 1)}} x_2.
\end{equation*}
From \eqref{eq:cond}, there exists $\epsilon_1 > 0$ such that
\begin{equation}
    \frac{\lambda_1}{\sqrt{2 \alpha (\lambda_2 +1)}} = \frac{(\alpha + 1) \lambda_2 + \alpha - 1}{\alpha (\lambda_2 + 1)} -\epsilon_1.
\end{equation}
Since \eqref{eq:x1eta1} implies $x_2^2 > 2 \alpha (\lambda_2 +1) L (V -N)$, this yields
$
    \dot V = \dot W_1 \le -\epsilon_1 x_2 \le - \epsilon_1 \sqrt{2 \alpha (\lambda_2 +1) L} \sqrt{V-N}.
    $
    In the second case in \eqref{eq:V}, one has $\dot V = \dot W_2$, and from $x_1 \changed{\ge} \frac{x_2^2}{4 \alpha (\lambda_2 +1)L} = V > N$, one obtains $x_1 - \eta > 0$ and
\begin{align}
    \dot W_2 &= \frac{x_2}{2 \alpha (\lambda_2 + 1)L}(-\lambda_2 L \spow{x_1 -\eta}{0} - \ddot f) \nonumber \\
    &\le - \frac{\lambda_2 - 1}{2 \alpha (\lambda_2 + 1)} x_2.
\end{align}
Since $x_2 = 2 \sqrt{\alpha (\lambda_2+1) L V}$, this implies 
\begin{equation}
    \dot V = \dot W_2 \le - \frac{(\lambda_2 - 1)\sqrt{L}}{\sqrt{\alpha (\lambda_2 + 1)}} \sqrt{V-N}.
\end{equation}
In the third case in \eqref{eq:V}, one has $\dot V = \dot W_3$, with
\begin{align}
    \dot W_3 &= -\lambda_1 \sqrt{L} \spowf{x_1 - \eta}{1}{2} + x_2 \nonumber \\
    &\quad - \frac{x_2}{(\lambda_2 + 1) L} ( -\lambda_2 L \spow{x_1 - \eta}{0} - \ddot f) \nonumber \\
    &\le -\lambda_1 \sqrt{L} \spowf{x_1 - \eta}{1}{2} + 2 x_2.
\end{align}
From \eqref{eq:cond}, there exists $\epsilon_2 > 0$ such that $\lambda_1$ may be written as $\lambda_1 = 2 \sqrt{2 (\lambda_2 + 1)} + \epsilon_2$.
Moreover, 
\begin{equation}
    x_1 - \eta \ge \frac{x_2^2}{2 (\lambda_2 + 1) L} + V - N
\end{equation}
by definition of $V$, and consequently $x_1 - \eta \ge \frac{x_2^2}{2 (\lambda_2+1) L}$ as well as $x_1 - \eta \ge V - N$.
Hence,
\begin{align}
    \dot V &= \dot W_3 \le - \epsilon_2 \sqrt{L} \sqrt{V - N} \nonumber \\
    &\quad- 2 \sqrt{2 (\lambda_2 + 1) L} \sqrt{x_1 - \eta} + 2x_2 \le - \epsilon_2 \sqrt{L} \sqrt{V - N}.
\end{align}
\changed{Finally, computing Clarke's generalized gradient and applying \cite[Lemma 8]{polyakov2014stability} yields $\dot V \in \operatorname{co} \{ \dot W_1, \dot W_2 \}$ or $\dot V \in \operatorname{co} \{ \dot W_2, \dot W_3 \}$ on the respective borders of the regions in \eqref{eq:W}, concluding the proof}.\qed

\subsection{Proof of Lemma~\ref{lem:inv}}
\label{app:proofs:inv}

Applying the comparison lemma \cite[Lemma 3.4]{khalil2002nonlinear} to \eqref{eq:Vdot} proves existence of a finite time $\tau$, depending only on $V(\x(0))$, such that $V(\x(t))\le N$ for $t \ge \tau$.
To see also \eqref{eq:supV}, note that this is straightforwardly obtained in the second case in \eqref{eq:W}, while in the first case
$
    \frac{x_2^2}{4 \alpha (\lambda_2 + 1) L} \ge x_1 = \frac{x_2^2}{2 \alpha (\lambda_2 + 1) L} - V
    $,
    and in the third case
$
    \frac{(2\alpha + 1) x_2^2}{4 \alpha (\lambda_2 +1) L} \le x_1 = \frac{x_2^2}{2 (\lambda_2 + 1) L} + V
    $
    each imply the inequality $\frac{x_2^2}{4 \alpha (\lambda_2 + 1) L} \le N$ due to $V \le N$.\qed

\endgroup

\bibliographystyle{elsarticle-harv}
\bibliography{literature}           

\begin{thebibliography}{20}
\expandafter\ifx\csname natexlab\endcsname\relax\def\natexlab#1{#1}\fi
\providecommand{\url}[1]{\texttt{#1}}
\providecommand{\href}[2]{#2}
\providecommand{\path}[1]{#1}
\providecommand{\DOIprefix}{doi:}
\providecommand{\ArXivprefix}{arXiv:}
\providecommand{\URLprefix}{URL: }
\providecommand{\Pubmedprefix}{pmid:}
\providecommand{\doi}[1]{\href{http://dx.doi.org/#1}{\path{#1}}}
\providecommand{\Pubmed}[1]{\href{pmid:#1}{\path{#1}}}
\providecommand{\bibinfo}[2]{#2}
\ifx\xfnm\relax \def\xfnm[#1]{\unskip,\space#1}\fi
\bibitem[{Angulo et~al.(2012a)Angulo, Moreno and
  Fridman}]{angulo2012differentiation}
\bibinfo{author}{Angulo, M.T.}, \bibinfo{author}{Moreno, J.A.},
  \bibinfo{author}{Fridman, L.}, \bibinfo{year}{2012}a.
\newblock \bibinfo{title}{The differentiation error of noisy signals using the
  generalized super-twisting differentiator}, in: \bibinfo{booktitle}{Proc.
  51st IEEE Conf. on Decision and Control}, pp. \bibinfo{pages}{7383--7388}.
\bibitem[{Angulo et~al.(2012b)Angulo, Moreno and Fridman}]{angulo2012optimal}
\bibinfo{author}{Angulo, M.T.}, \bibinfo{author}{Moreno, J.A.},
  \bibinfo{author}{Fridman, L.}, \bibinfo{year}{2012}b.
\newblock \bibinfo{title}{Optimal gain for the super-twisting differentiator in
  the presence of measurement noise}, in: \bibinfo{booktitle}{2012 American
  Control Conference (ACC)}, pp. \bibinfo{pages}{6154--6159}.
\bibitem[{Cruz-Zavala and Moreno(2019)}]{cruz2019levant}
\bibinfo{author}{Cruz-Zavala, E.}, \bibinfo{author}{Moreno, J.A.},
  \bibinfo{year}{2019}.
\newblock \bibinfo{title}{Levant's arbitrary-order exact differentiator: a
  {Lyapunov} approach}.
\newblock \bibinfo{journal}{IEEE Trans. on Aut. Control} \bibinfo{volume}{64},
  \bibinfo{pages}{3034--3039}.
\bibitem[{Efimov and Fridman(2011)}]{efimov2011hybrid}
\bibinfo{author}{Efimov, D.V.}, \bibinfo{author}{Fridman, L.},
  \bibinfo{year}{2011}.
\newblock \bibinfo{title}{A hybrid robust non-homogeneous finite-time
  differentiator}.
\newblock \bibinfo{journal}{IEEE Trans. on Aut. Control} \bibinfo{volume}{56},
  \bibinfo{pages}{1213--1219}.
\bibitem[{Filippov(1988)}]{filippov1988differential}
\bibinfo{author}{Filippov, A.F.}, \bibinfo{year}{1988}.
\newblock \bibinfo{title}{Differential Equations with Discontinuous Right-Hand
  Side}.
\newblock \bibinfo{publisher}{Kluwer Academic Publishing},
  \bibinfo{address}{Dortrecht, The Netherlands}.
\bibitem[{Khalil(2002)}]{khalil2002nonlinear}
\bibinfo{author}{Khalil, H.K.}, \bibinfo{year}{2002}.
\newblock \bibinfo{title}{Nonlinear Systems}.
\newblock \bibinfo{edition}{Third} ed., \bibinfo{publisher}{Prentice Hall},
  \bibinfo{address}{Upper Saddle River, NJ, USA}.
\bibitem[{Khalil and Praly(2014)}]{khapra_ijrnc14}
\bibinfo{author}{Khalil, H.K.}, \bibinfo{author}{Praly, L.},
  \bibinfo{year}{2014}.
\newblock \bibinfo{title}{High-gain observers in nonlinear feedback control}.
\newblock \bibinfo{journal}{Int. Journal of Robust and Nonlinear Control}
  \bibinfo{volume}{24}, \bibinfo{pages}{993--1015}.
\bibitem[{Levant(1998)}]{levant1998robust}
\bibinfo{author}{Levant, A.}, \bibinfo{year}{1998}.
\newblock \bibinfo{title}{Robust exact differentiation via sliding mode
  technique}.
\newblock \bibinfo{journal}{Automatica} \bibinfo{volume}{34},
  \bibinfo{pages}{379--384}.
\bibitem[{Levant(2003)}]{levant_ijc03}
\bibinfo{author}{Levant, A.}, \bibinfo{year}{2003}.
\newblock \bibinfo{title}{Higher-order sliding modes, differentation and
  output-feedback control}.
\newblock \bibinfo{journal}{International Journal of Control}
  \bibinfo{volume}{76}, \bibinfo{pages}{924--941}.
\bibitem[{Levant and Livne(2020)}]{levliv_ejc20}
\bibinfo{author}{Levant, A.}, \bibinfo{author}{Livne, M.},
  \bibinfo{year}{2020}.
\newblock \bibinfo{title}{Robust exact filtering differentiators}.
\newblock \bibinfo{journal}{European Journal of Control} \bibinfo{volume}{55},
  \bibinfo{pages}{33--44}.
\bibitem[{Levant et~al.(2017)Levant, Livne and Yu}]{levant2017sliding}
\bibinfo{author}{Levant, A.}, \bibinfo{author}{Livne, M.}, \bibinfo{author}{Yu,
  X.}, \bibinfo{year}{2017}.
\newblock \bibinfo{title}{Sliding-mode-based differentiation and its
  application}, in: \bibinfo{booktitle}{20th IFAC World Congress},
  \bibinfo{publisher}{Elsevier}. pp. \bibinfo{pages}{1699--1704}.
\bibitem[{Mboup et~al.(2009)Mboup, Join and Fliess}]{mbojoi_na09}
\bibinfo{author}{Mboup, M.}, \bibinfo{author}{Join, C.},
  \bibinfo{author}{Fliess, M.}, \bibinfo{year}{2009}.
\newblock \bibinfo{title}{Numerical differentiation with annihilators in noisy
  environment}.
\newblock \bibinfo{journal}{Numerical Algorithms} \bibinfo{volume}{50},
  \bibinfo{pages}{439--467}.
\bibitem[{Mboup and Riachy(2018)}]{mboup2018frequency}
\bibinfo{author}{Mboup, M.}, \bibinfo{author}{Riachy, S.},
  \bibinfo{year}{2018}.
\newblock \bibinfo{title}{Frequency-domain analysis and tuning of the algebraic
  differentiators}.
\newblock \bibinfo{journal}{International Journal of Control}
  \bibinfo{volume}{91}, \bibinfo{pages}{2073--2081}.
\bibitem[{Mojallizadeh et~al.(2021)Mojallizadeh, Brogliato and
  Acary}]{mojallizadeh2021time}
\bibinfo{author}{Mojallizadeh, M.R.}, \bibinfo{author}{Brogliato, B.},
  \bibinfo{author}{Acary, V.}, \bibinfo{year}{2021}.
\newblock \bibinfo{title}{Time-discretizations of differentiators: Design of
  implicit algorithms and comparative analysis}.
\newblock \bibinfo{journal}{Int. Journal of Robust and Nonlinear Control}
  \bibinfo{volume}{31}, \bibinfo{pages}{7679--7723}.
\bibitem[{Moreno(2022)}]{moreno2022arbitrary}
\bibinfo{author}{Moreno, J.A.}, \bibinfo{year}{2022}.
\newblock \bibinfo{title}{Arbitrary-order fixed-time differentiators}.
\newblock \bibinfo{journal}{IEEE Trans. on Aut. Control} \bibinfo{volume}{67},
  \bibinfo{pages}{1543--1549}.
\bibitem[{Polyakov and Fridman(2014)}]{polyakov2014stability}
\bibinfo{author}{Polyakov, A.}, \bibinfo{author}{Fridman, L.},
  \bibinfo{year}{2014}.
\newblock \bibinfo{title}{Stability notions and lyapunov functions for sliding
  mode control systems}.
\newblock \bibinfo{journal}{Journal of the Franklin Institute}
  \bibinfo{volume}{351}, \bibinfo{pages}{1831--1865}.
\bibitem[{Reichhartinger et~al.(2018)Reichhartinger, Efimov and
  Fridman}]{reichhartinger2018special}
\bibinfo{author}{Reichhartinger, M.}, \bibinfo{author}{Efimov, D.},
  \bibinfo{author}{Fridman, L.}, \bibinfo{year}{2018}.
\newblock \bibinfo{title}{Special issue on differentiators}.
\newblock \bibinfo{journal}{International Journal of Control}
  \bibinfo{volume}{91}, \bibinfo{pages}{1980--1982}.
\bibitem[{Seeber(2020)}]{seeber2020computing}
\bibinfo{author}{Seeber, R.}, \bibinfo{year}{2020}.
\newblock \bibinfo{title}{Computing and estimating the reaching time of the
  super-twisting algorithm}, in: \bibinfo{booktitle}{Variable-Structure Systems
  and Sliding-Mode Control}. \bibinfo{publisher}{Springer},
  \bibinfo{address}{Cham, Switzerland}, pp. \bibinfo{pages}{73--123}.
\bibitem[{Seeber and Haimovich(2023)}]{seeber2023optimal}
\bibinfo{author}{Seeber, R.}, \bibinfo{author}{Haimovich, H.},
  \bibinfo{year}{2023}.
\newblock \bibinfo{title}{Optimal robust exact differentiation via linear
  adaptive techniques}.
\newblock \bibinfo{journal}{Automatica} \bibinfo{volume}{148},
  \bibinfo{pages}{110725}.
\bibitem[{Vasiljevic and Khalil(2008)}]{vasiljevic2008error}
\bibinfo{author}{Vasiljevic, L.K.}, \bibinfo{author}{Khalil, H.K.},
  \bibinfo{year}{2008}.
\newblock \bibinfo{title}{Error bounds in differentiation of noisy signals by
  high-gain observers}.
\newblock \bibinfo{journal}{Systems \& Control Letters} \bibinfo{volume}{57},
  \bibinfo{pages}{856--862}.

\end{thebibliography}

\end{document}